\documentclass[aps,twocolumn,showpacs,preprintnumbers,superscriptaddress,nofootinbib]{revtex4}
\usepackage{subfigure,graphicx,epsfig,amsmath,amsfonts,amssymb,xcolor,slashed,ulem}

\usepackage{float}
\usepackage{wrapfig}

\newcommand{\minv}{M_{\rm inv}}

\newcommand{\br}{{\rm Br}}
\renewcommand\sout{\bgroup \color[rgb]{1,0,0} \ULdepth=-.5ex \ULset}
\allowdisplaybreaks[1]
\usepackage{graphicx}
\graphicspath{{Images/}}
\usepackage{caption}
\usepackage[font={small}]{caption}
\DeclareMathSizes{12}{12}{12}{12}

\usepackage{amsmath,amssymb,amsfonts}	
\usepackage{color}
\usepackage{slashed}            
\usepackage{bbm}                
\usepackage{xspace}				

\numberwithin{equation}{section} 
\usepackage{tikz}
\usetikzlibrary{arrows,shapes}
\usetikzlibrary{trees}
\usetikzlibrary{matrix,arrows} 				
\usetikzlibrary{positioning}				
\usetikzlibrary{calc,through}				
\usetikzlibrary{decorations.pathreplacing}  
\usepackage{pgffor}							

\usetikzlibrary{decorations.pathmorphing}	
\usetikzlibrary{decorations.markings}
\tikzset{
vector/.style={decorate, decoration={snake}, draw},
provector/.style={decorate, decoration={snake,amplitude=2.5pt}, draw},
antivector/.style={decorate, decoration={snake,amplitude=-2.5pt}, draw},
fermion/.style={draw=black, postaction={decorate},
	decoration={markings,mark=at position .55 with {\arrow[draw=black]{>}}}},
fermionbar/.style={draw=black, postaction={decorate},
	decoration={markings,mark=at position .55 with {\arrow[draw=black]{<}}}},
fermionnoarrow/.style={draw=black},
gluon/.style={decorate, draw=black,
	decoration={coil,amplitude=4pt, segment length=5pt}},
scalar/.style={dashed,draw=black, postaction={decorate},
	decoration={markings,mark=at position .55 with {\arrow[draw=black]{>}}}},
scalarbar/.style={dashed,draw=black, postaction={decorate},
	decoration={markings,mark=at position .55 with {\arrow[draw=black]{<}}}},
scalarnoarrow/.style={dashed,draw=black},
electron/.style={draw=black, postaction={decorate},
	decoration={markings,mark=at position .55 with {\arrow[draw=black]{>}}}},
bigvector/.style={decorate, decoration={snake,amplitude=4pt}, draw},
}

\tikzstyle{block} = [draw, rectangle, 
    minimum height=3em, minimum width=6em]

\begin{document}
\title{Triangle singularities in $B^-\rightarrow D^{*0}\pi^-\pi^0\eta$ and
$B^-\rightarrow D^{*0}\pi^-\pi^+\pi^-$}
\date{\today}

\author{R. Pavao}
\email{rpavao@ific.uv.es}

\author{S. Sakai}

\author{E.~Oset}
\affiliation{Departamento de
F\'{\i}sica Te\'orica and IFIC, Centro Mixto Universidad de
Valencia-CSIC Institutos de Investigaci\'on de Paterna, Aptdo.22085,
46071 Valencia, Spain}

\begin{abstract} 
 The possible role of the triangle mechanism in the $B^-$ decay into
 $D^{*0}\pi^-\pi^0\eta$ and $D^{*0}\pi^-\pi^+\pi^-$ is investigated.
 In this process, the triangle singularity appears from the decay of
 $B^-$ into $D^{*0}K^-K^{*0}$ followed by the decay of
 $K^{*0}$ into $\pi^-K^+$ and the fusion of the $K^+K^-$ which
 {forms} the $a_0(980)$ or $f_0(980)$ {which} finally {decay} into $\pi^0\eta$ or
 $\pi^+\pi^-$ respectively.
 The triangle mechanism from the $\bar{K}^*K\bar{K}$ loop generates a
 peak around 1420 MeV in the invariant mass of $\pi^-a_0$ or $\pi^-f_0$,
 and gives sizable branching {fractions} $\br(B^-\rightarrow
 D^{*0}\pi^- a_0;a_0 \rightarrow \pi^0\eta)= 1.66 \times 10^{-6}$ and $\br(B^-\rightarrow
 D^{*0}\pi^- f_0 ; f_0 \rightarrow \pi^+\pi^-)= 2.82 \times 10^{-6}$.
\end{abstract}

\maketitle
\section{Introduction}
Hadron spectroscopy is a way to investigate Quantum
Chromodynamics (QCD), which is the basic theory of the strong
interaction.
The success of the quark model in the low-lying
hadron spectrum gives us an interpretation of the
baryons as a composite of three quarks, and the mesons as that of
quark and anti-quark~\cite{Godfrey:1985xj,Capstick:1986bm}.
Meanwhile, the possibility of non conventional hadrons called exotics,
which are not prohibited by QCD, have been intensively studied.
One example is the $\Lambda(1405)$:
the quark model predicts a mass at higher energy than the observed peak, and
a $\bar{K}N$ $(I=0)$ molecular state seems to give a better description
as originally studied in Ref.~\cite{Dalitz:1967fp} followed by many
studies which are summarized in
Refs.~\cite{Hyodo:2011ur,Kamiya:2016jqc}.
The spectrum of the low-lying scalar mesons, such as $f_0(980)$ or $a_0(980)$
mesons, is also discussed in this picture
\cite{Weinstein:1982gc,Weinstein:1983gd,Weinstein:1990gu}, {while
the possible explanation as tetraquark states is also discussed in
Refs.}~\cite{Jaffe:1976ig,Jaffe:1976ih}.
These days, in the heavy sector, the $XYZ$~\cite{pdg} {and} the
$P_c$~\cite{Aaij:2015tga,Aaij:2015fea} {were discovered},
which cannot be associated with the states predicted {by} the quark
model. Another sort of non conventional hadrons are the molecular states of other hadrons, which have been often invoked to describe many existing states (see recent review in Ref.~\cite{fkguo}).
Besides ordinary hadrons, molecular states or multiquark states, triangle singularities can generate peaks, but these peaks
appear from a simple kinematical effect.
These singularities were pointed out by Landau~\cite{Landau:1959fi}, and the
Coleman-Norton theorem says that the singularity appears when the process
has a classical counterpart~\cite{Coleman:1965xm}:
in the decay process of a particle $1$ into the particles $2$ and
$3$, the particle $1$ decays first into particles $A$ and $B$, followed by the decay of
$A$ into the particles $2$ and $C$, and finally the particles $BC$ merge
into the particle $3$.
The particles $A$, $B$, and $C$ are the intermediate particles,
and the singularity appears if the momenta of these intermediate particles
can take on-shell values.
A novel way to understand this process
is proposed in Ref.~\cite{Bayar:2016ftu}.

For the decay of $\eta(1405)$ into $\pi^0\pi^0\eta$ via $\pi^0a_0$ and
$\pi^0\pi^+\pi^-$ via $\pi^0f_0$, the triangle mechanism gives a good
explanation~\cite{Wu:2011yx,Aceti:2012dj,Wu:2012pg}.
The $K^*\bar{K}K$ loop generates the triangle singularity in this process, and
the anomalously large branching fraction of the isospin-violating
$\pi^0f_0$ channel reported by BESIII~\cite{BESIII:2012aa} is well
explained with the mechanism.

The peak associated with this singularity can be misidentified with {a}
resonance state.
For example, the studies in
Refs.~\cite{Liu:2015taa,Wang:2013cya,Liu:2013vfa} suggest the possible
explanation of $Z_c(3900)$ with the triangle
mechanism. Similarly, a peak seen in the $\pi f_0 (980)$ mass distribution, identified as the "$a_1(1420)$" meson by the COMPASS colaboration \cite{Adolph:2015pws}, is shown to be a
manifestation of the triangle mechanism as studied in
Refs.~\cite{Liu:2015taa,Ketzer:2015tqa,Aceti:2016yeb}.
In particular, many $XYZ$ states are discovered as a peak of the
invariant mass distribution in the heavy hadron decay.
Then, the thorough study on the role of the triangle
singularities in the heavy hadron decay is important to
clarify the properties of the reported $XYZ$ states.
In the $B^-\rightarrow K^-\pi^-D_{s0}^+(2317)$
$(K^-\pi^-D_{s1}^+(2460))$ process, a peak can be generated by the
triangle mechanism around 2800 MeV (2950 MeV) in the $\pi^-D_{s0}^+$
($\pi^-D_{s1}^+$) invariant mass spectrum,
which is driven by the $K^*DK$ $(K^*D^*K)$ loop, and gives a sizable
branching fraction into the channel~\cite{Sakai:2017hpg}.
The $D_{s0}^+$ and $D_{s1}^+$ in the final state are
dynamically generated by the $DK$ and $D^*K$, and have large coupling
with these states~\cite{Gamermann:2006nm,Gamermann:2007fi,Torres:2014vna}.
Because the process of the triangle mechanism contains a fusion of two
hadrons, the existence of a hadronic molecular state plays {an} important
role {in having} a measurable strength.
Then, the study of the singularity is also a useful tool to study the
hadronic molecular {states}.
{Regarding the} $P_c$ {peak,} discovered in the $J/\psi p$ invariant mass
distribution of the $\Lambda_b$ decay~\cite{Aaij:2015tga,Aaij:2015fea},
the possibility of the interpretation as a triangle singularity
was pointed out in Refs.~\cite{Guo:2015umn,Oller:1997ti}. However, in Ref.~\cite{Bayar:2016ftu} it was noted that if the $P_c$ quantum numbers were $\frac{1}{2}^+$ or $\frac{3}{2}^+$ the triangle mechanism could provide an interpretation of the narrow experimental peak, but not if the quantum numbers are $\frac{3}{2}^-$, $\frac{5}{2}^+$, as preferred by experiment.

\begin{figure}[t!]
\begin{tikzpicture}[line width=1.5 pt]
	\draw[fermion] (-1.5,2) -- (1,2);
	\draw[fermion] (1,2) -- (4,2);
	\draw[fermion] (4,2) -- (6,2);
	\draw[fermion] (4,2) -- (3,0);
	\draw[fermion] (1,2) -- (3,0);
	%
	\draw[fermion] (1,2) -- (1.5,4);
	\draw[fermionnoarrow] (3.03,0.07) -- (4.1,-1);
	\draw[fermionnoarrow] (3,0) -- (4.05,-1.05);
	
    \node at (-1.3, 2.4) {$B^-$};
    \node at (0.5,3.3) {$D^{* 0}$};
    
    \node at (2, 2.4) {$K^{* 0}$};
    \node at (2.9, 2.35) {$(P-q)$};    
    
    \node at (4, 1) {$K^+$};
    \node at (5.1,0.93) {$(P-q-k)$};    
    
    \node at (0.6, 1) {$K^-$};
    \node at (1.1, 0.95) {$(q)$};    
    
    \node at (5, 2.4) {$\pi^-$};
    \node at (5.5, 2.3) {$(k)$};    
    
    \node at (3.9, -.4) {$R$};
\end{tikzpicture}
  \caption{Diagram for the decay of $B^-$ into $D^{* 0}$, $\pi^ -$ and $R$, where $R=a_0(980) \ \text{or} \ f_0(980)$.}
  \label{fig::diagramBtoDpR}
\end{figure}
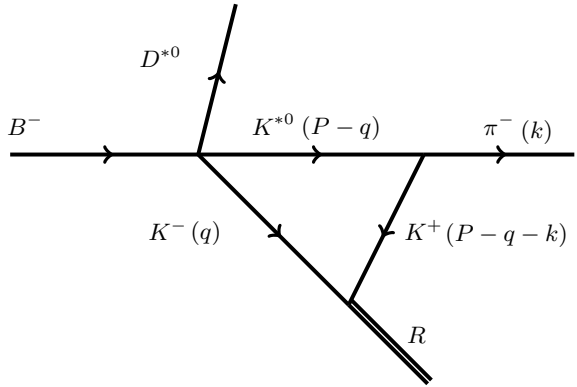

In the present study, we investigate the $B^-\rightarrow
D^{*0}\pi^-\pi^0\eta$ and $B^-\rightarrow D^{*0}\pi^-\pi^+\pi^-$ decays
via $a_0$ and $f_0$ formation.
The process of $B^-\rightarrow D^{*0}K^-K^{*0}$ followed by the
$K^{*0}$ decay into $\pi^-K^+$ and the merging of the $K^+K^-$ into $a_0$
or $f_0$ {(see Fig.} \ref{fig::diagramBtoDpR}{)} generate a singularity, which would appear around
$1418$ MeV in the invariant mass of $\pi^-a_0$ or $\pi^-f_0$, {as calculated using Eq.} (18) of Ref.~\cite{Bayar:2016ftu}.
In this study, these $a_0$ and $f_0$ states appear as the
dynamically generated {states} of $\pi\pi$, $K\bar{K}$, $\eta\eta$, and
 $K\bar{K}$, $\pi^0\eta$ in the $I=0$ and $I=1$ channels, respectively,
as studied in Refs.~\cite{Oller:1997ti,Oller:1998hw}.

The mechanism proposed here, without the indication of how the $K^* \bar{K}$ could be formed, and without a quantitative evaluation of the process, was suggested in Ref.~\cite{Liu:2015taa}. We provide here a realistic example of a physical process where this can occur, which also allows us to perform a quantitative calculation of the amplitudes involved.

Weak decays of heavy hadrons are turning into a good laboratory to find many triangle singularities. Apart from the work of Ref.~\cite{missingref}, the $B_c \rightarrow B_s \pi \pi$ reaction has been suggested, where $B^+_c \rightarrow \bar{K}^{* 0} B^+$, $\bar{K}^{* 0} \rightarrow\pi^0 \bar{K}^0$ and $\bar{K}^0 B^+ \rightarrow \pi B^0_s$. Yet, there are large uncertainties quantizing the $\bar{K}^0 B^+ \rightarrow \pi B^0_s$ amplitude.

In the present case we rely upon well known $K\bar{K} \rightarrow \pi \pi$ ($K\bar{K} \rightarrow \pi \eta$) amplitudes, and the $B^- \rightarrow D^{* 0} \bar{K}^{* 0} K^-$ vertex can be obtained from experiment. Hence, we are able to quantize the decay rates of the mechanism proposed and we find that the mass distribution of these decay processes shows a
peak associated with the triangle singularity, and finally find the branching {fractions} $\br(B^-\rightarrow
 D^{*0}\pi^- a_0; a_0 \rightarrow \pi^0\eta)= 1.66 \times 10^{-6}$ and $\br(B^-\rightarrow
 D^{*0}\pi^- f_0; f_0 \rightarrow \pi^+\pi^-)= 2.82 \times 10^{-6}$.
 
\section{Formalism}

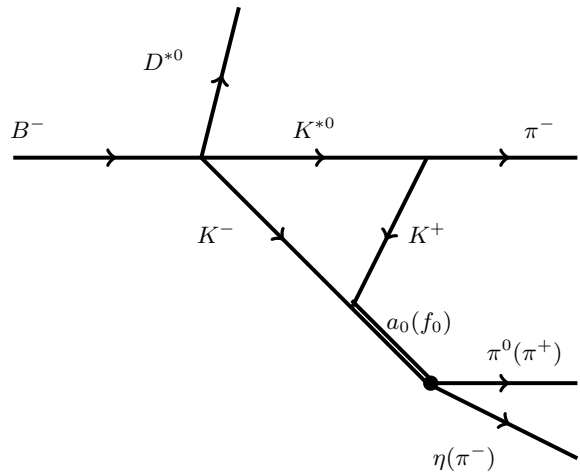
\begin{figure}[h!]
\begin{tikzpicture}[line width=1.5 pt]
	\draw[fermion] (-1.5,2) -- (1,2);
	\draw[fermion] (1,2) -- (4,2);
	\draw[fermion] (4,2) -- (6,2);
	\draw[fermion] (4,2) -- (3,0);
	\draw[fermion] (1,2) -- (3,0);

	\draw[fermion] (1,2) -- (1.5,4);
	\draw[fermionnoarrow] (3,0.1) -- (4.1,-1);
	\draw[fermionnoarrow] (3,0) -- (4.05,-1.05);
	\draw[fermion] (4,-1) -- (6,-1);
	\draw[fermion] (4,-1) -- (6,-2);
	\draw[fill=black] (4.05,-1) circle (.075cm);

    \node at (-1.3, 2.4) {$B^-$};
    \node at (0.5,3.3) {$D^{* 0}$};
    \node at (2.5, 2.4) {$K^{* 0}$};
    \node at (4, 1) {$K^+$};
    \node at (1.2, 1) {$K^-$};
    \node at (5.5, 2.4) {$\pi^-$};
    \node at (3.9, -.2) {$a_0(f_0)$};
    \node at (5.3, -0.6) {$\pi^0 (\pi^+)$};
    \node at (4.5, -2) {$\eta (\pi^-)$};
\end{tikzpicture}
  \caption{Diagram for the decay of $B^- \rightarrow D^{* 0} \pi^- \eta \pi^0 (\pi^+ \pi^-)$.}
  \label{fig::totaldecay}
\end{figure}

We will analyse the effect of triangle singularities in the following decays: $B^- \rightarrow D^{* 0} \pi^- \eta \pi^0$ and $B^- \rightarrow D^{* 0} \pi^- \pi^+ \pi^-$. The complete Feynman diagram for these decays, with the triangle mechanism through the $a_0$ or $f_0$ mesons, is shown in Fig. \ref{fig::totaldecay}.

At first, we evaluate the $B^- \rightarrow D^{*} \pi R \ (R=a_0,f_0)$. This then produces the triangle diagram shown in Fig. \ref{fig::diagramBtoDpR}. The $T$ matrix $t_{B \rightarrow D^* \pi R}$ will have the following form,
\begin{widetext}
\begin{equation}
\label{eq:t_total0}
-i t_{B \rightarrow D^* \pi R} = i \sum_{\text{pol. of } K^* } \int \frac{d^4 q}{(2 \pi)^4}  \frac{i \ t_{B^- \rightarrow D^{* 0} K^{* 0} K^-}}{q^2-m_K^2+i \epsilon} \frac{i \ t_{K^*K^+ \pi^-} }{(P-q)^2-m_{K^*}^2+i \epsilon} \frac{i \  t_{K^+K^- , R}}{(P-q-k)^2 - m_K^2+i\epsilon}.
\end{equation}
\end{widetext}
The amplitude in Eq. \eqref{eq:t_total0} is evaluated in the center-of-mass (CM) frame of $\pi R$. Now we need to calculate the three vertices, $t_{B^- \rightarrow D^{* 0} K^{* 0} K^-} $, $t_{K^*K^+ \pi^-}$ and $t_{K^+K^- , R}$, in Eq. \eqref{eq:t_total0}.

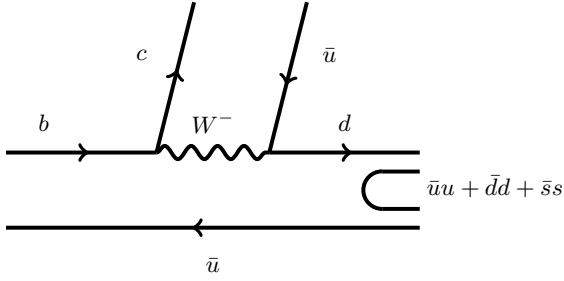
\begin{figure}[t!]
%
\begin{tikzpicture}[line width=1.5 pt]
	\draw[fermion] (-1.5,2) -- (0.5,2);
	\draw[vector] (0.5,2) -- (2,2);
	\draw[fermion] (2,2) -- (4,2);
	\draw[fermion] (0.5,2) -- (1,4);
	\draw[fermion] (2.5,4) -- (2,2);
	\draw[fermion] (4,1) -- (-1.5,1);
	\draw[fermionnoarrow] (3.5,1.25) arc (270:90:.25);
	\draw[fermionnoarrow] (3.5,1.747) -- (4,1.747);
	\draw[fermionnoarrow] (3.5,1.25) -- (4,1.25);
    \node at (-1, 2.4) {$b$};
    \node at (0.3,3.3) {$c$};
    \node at (1.25, 2.4) {$W^{-}$};
    \node at (3, 2.4) {$d$};
    \node at (2.8, 3.3) {$\bar{u}$};
    \node at (1.25, 0.5) {$\bar{u}$};
    \node at (5, 1.55) {$\bar{u}u+\bar{d}d+\bar{s}s$};
\end{tikzpicture}
 \caption{Diagram for the decay of $B^-$ into $D^{* 0}$, $K^{* 0}$ and $K^-$ as seen through the quark constituents of the hadrons.}
  \label{fig::internal}
\end{figure}

First, we discuss the $B^-\rightarrow D^{*0}K^-K^{*0}$ vertex.
At the quark level, the Cabibbo-allowed vertex is formed
 through an internal emission of a $W$
boson \cite{Chau:1982da} (as can be seen in Fig. \ref{fig::internal}), producing a $c \bar{u}$ that forms the $D^{* 0}$, with the remaining $d \bar{u}$ quarks hadronizing and producing the $K^-$ and $K^{* 0}$ mesons
with the selection of the $\bar{s}s$ pair from a created vacuum $\bar{u}u+\bar{d}d+\bar{s}s$ state.
Since both $D^{* 0}$ and $K^{* 0}$ have $J^P=1^-$, the interaction in the $B^- \rightarrow D^{* 0} K^- K^{* 0}$ vertex can proceed via $s$-wave and we take the amplitude of the form,
\begin{equation}
\label{eq:t_1}
 t_{B^- \rightarrow D^{* 0} K^{* 0} K^-} = C \ \epsilon_{\mu}(K^*) \epsilon^{\mu}(D^*).
\end{equation}
Given that we know that the branching ratio of this decay is $\br(B^- \rightarrow D^{* 0} K^{* 0} K^-)=1.5 \times 10^{-3}$~\cite{pdg,Drutskoy:2002ib}, we can determine the constant $C$ by calculating the width of this decay,
\begin{widetext}
\begin{equation}
\label{eq:decayBtoDKK}
\frac{d \Gamma_{B^- \rightarrow D^{* 0} K^{* 0} K^-}}{d M_{\text{inv}}(K^* D^*)}=\frac{1}{(2 \pi)^3} \frac{|\vec{p}_{K^-}| |\vec{\tilde{p}}_{K^*}|}{4 M_B^2} \overline{\sum} \sum \left|t_{B^- \rightarrow D^{* 0} K^{* 0} K^-} \right|^2,
\end{equation}
\end{widetext}
where $\vec{p}_{K^-}$ is the momentum of $K^-$ in the $B^-$ rest frame, and $\vec{\tilde{p}}_{K^*}$ is the momentum of $K^{*0}$ in the $K^{* 0} D^{* 0}$ CM frame. The absolute values of both momenta are given by
\begin{subequations}
\begin{align}
&	|\vec{p}_{K^-}| = \frac{\lambda^{1/2} (M_B^2, m^2_{K^-}, M^2_{\text{inv}}(K^* D^*))}{2 M_B},	\\
&   |\vec{\tilde{p}}_{K^*}|=\frac{\lambda^{1/2} (M^2_{\text{inv}}(K^* D^*), m^2_{K^*}, m^2_{D^*})}{2 M_{\text{inv}}(K^* D^*)},
\end{align}
\end{subequations}
with $\lambda(x,y,z)$ the ordinary K{\"a}llen function.

Now, if we square the $T$ matrix in \eqref{eq:t_1} and sum over the polarizations, we get
\begin{align}
\overline{\sum} \sum &\left|t_{B^- \rightarrow D^{* 0} K^{* 0} K^-} \right|^2\\
=&C^2 \sum_{\text{pol}} \epsilon_{\mu}(K^*)\epsilon_{\nu}(K^*) \epsilon^{\mu}(D^*) \epsilon^{\nu}(D^*)
\label{eq:t_1avg1}\\
=&C^2 \left(2+ \frac{(p_{K^*} \cdot p_{D^*})^2}{m^2_{K^*}
 m^2_{D^*}}\right)\\
 =&C^2 \left(2+ \frac{\left(M_{\text{inv}}^2(K^* D^*)-m^2_{K^*}-m^2_{D^*}\right)^2}{4 m^2_{K^*} m^2_{D^*}}\right). 
\end{align}
where we used the fact that $(p_{K^*}+p_{D^*})^2=M_{\text{inv}}^2(K^* D^*)$, $i.e.$, $p_{K^*} \cdot p_{D^*}= \frac{1}{2} \left(M_{\text{inv}}^2(K^* D^*)-m^2_{K^*}-m^2_{D^*}\right)$.\\
Then, using this last equation in Eq.~\eqref{eq:decayBtoDKK}, we get
\begin{widetext} 
\begin{equation}
\label{eq:const}
\frac{C^2}{\Gamma_{B^-}}=\frac{\br(B^- \rightarrow D^{* 0} K^{* 0} K^-)}{\int dM_{\text{inv}}(K^* D^*) \frac{1}{(2 \pi)^3}  \frac{|\vec{p}_{K^-}| |\vec{\tilde{p}}_{K^*}|}{4 M_B^2}\left(2+ \frac{\left(M_{\text{inv}}^2(K^* D^*)-m^2_{K^*}-m^2_{D^*}\right)^2}{4 m^2_{K^*} m^2_{D^*}}\right)},
\end{equation}
\end{widetext} 
where the integral has the limits $M_{\text{inv}}(K^* D^*)|_{\text{min}}=m_{D^*}+m_{K^*}$ and $M_{\text{inv}}(K^* D^*)|_{\text{max}}=M_{B}-m_{K}$. 

Now we calculate the contribution of the vertex $K^{* 0} \rightarrow \pi^- K^+$. For that we will use the chiral invariant lagrangian with local hidden symmetry given in Refs.~\cite{LHS1,LHS2,LHS3,LHS4}, which is
\begin{equation}
\mathcal{L}_{VPP} = -i g \left < V^{\mu} \left[P, \partial_{\mu} P\right] \right >,
\end{equation}
where the $VPP$ subscript refers to the fact that we have a vertex with a vector and two pseudoscalar hadrons. The symbol $\left<...\right>$ stands for the trace over the $SU(3)$ flavour matrices, and $g=m_V/2 f_{\pi}$ is the coupling of the local hidden gauge, with $m_V=800 \ \text{MeV}$ and $f_{\pi}=93 \ \text{MeV}$.  The SU(3) matrices for the pseudoscalar and vector octet mesons $P$ and $V^{\mu}$ are given by
\begin{equation}
V_{\mu}=\begin{pmatrix}
\frac{1}{\sqrt{2}} \rho^0_{\mu} + \frac{1}{\sqrt{2}} \omega_{\mu} & \rho^+_{\mu} & K^{* +}_{\mu} \\ 
\rho^-_{\mu} & -\frac{1}{\sqrt{2}} \rho^0_{\mu} + \frac{1}{\sqrt{2}} \omega_{\mu} & K^{* 0}_{\mu} \\ 
K^{* -}_{\mu} & \bar{K}^{* 0}_{\mu} & \phi_{\mu}
\end{pmatrix},
\end{equation}
\begin{equation}
P=\begin{pmatrix}
\frac{1}{\sqrt{2}} \pi^0 + \frac{1}{\sqrt{6}} \eta  & \pi^+ &  K^+ \\
\pi^- & -\frac{1}{\sqrt{2}} \pi^0 + \frac{1}{\sqrt{6}} \eta & K^0\\
K^- & \bar{K}^0 & -\sqrt{\frac{2}{3}} \eta
\end{pmatrix}.
\end{equation}
Performing the matrix operations and the trace we get
\begin{equation}
\mathcal{L}_{K^*K\pi} =-ig {K^{* 0}}^{\mu} \left(K^- \partial_{\mu} \pi^+ - \pi^+ \partial_{\mu} K^- \right).
\end{equation}
So, for the $t$ matrix we get,
\begin{align}
\label{eq:thisone}
-i t_{K^*K^+ \pi^-} =& -i g \epsilon^{\mu}_{K^*}
 (p_{K^+}-p_{\pi})_\mu \\
 \simeq&  -i g \vec{\epsilon}_{K^*} \cdot (\vec{\tilde{p}}_{\pi}-\vec{\tilde{p}}_{K^+}),\label{eq_kskpi}
\end{align}
with $\vec{\tilde{p}}_{K^+}$ and $\vec{\tilde{p}}_{\pi}$ calculated in
the CM frame of $\pi R$. At the energy where the triangle singularity
appears ($M_{\text{inv}}(\pi R)=1418 \ \text{MeV}$), the momentum of $K^*$ is about $150 \
\text{MeV}/c$, which is small enough, compared with the mass of $K^*$,
to omit the zeroth component of the polarization vector in
Eq.~\eqref{eq:thisone}.

Finally we only need to calculate $t_{K^+K^- , R}$, before we can analyse the triangle diagram. The coupling of $R$ with $\pi^0 \eta$ or $\pi^+ \pi^-$ proceeds in $s$-wave. Then, the vertex is written simply as a constant,
\begin{equation}
t_{K^+K^- , R} = g_{K^- K^+ , R}.\label{eq_rkk}
\end{equation}
We can now analyse the effect of the triangle singularity on the $B^-
\rightarrow D^* \pi R$ decay.\\
Substituting Eqs.~\eqref{eq:t_1},~(\ref{eq_kskpi}) and (\ref{eq_rkk}) for
Eq.~(\ref{eq:t_total0}),
the decay amplitude $t_{B^- \rightarrow D^{* 0} \pi^- R}$
is  written as
\begin{widetext}
\begin{align}
\label{eq:t_total}
 t_{B^- \rightarrow D^* \pi R} = -i g_{K^- K^+, R}  \ g C \sum_{\text{pol. of $K^*$}} \int \frac{d^4 q}{(2 \pi)^4}  \  \frac{\vec{\epsilon}_{D^*} \cdot \vec{\epsilon}_{K^*}}{q^2-m_K^2+i \epsilon} \frac{\vec{\epsilon}_{K^*} \cdot (\vec{\tilde{p}}_{\pi}-\vec{\tilde{p}}_{K^+})}{(P-q)^2-m^2_{K^*}+i \epsilon} \frac{1}{(P-q-k)^2 - m_K^2+i \epsilon} \ ,
\end{align}
\end{widetext}
where for $t_{B^- \rightarrow D^{* 0} K^{* 0} K^-}$ we have also the spatial components of the polarization vectors, and $\vec{\tilde{p}}_{K^+}$, $\vec{\tilde{p}}_{K^+}$ are taken
in the CM frame of $\pi R$.
As we have mentioned below Eq.~(\ref{eq_kskpi}), the momentum of
the $K^{*0}$ around the triangle peak is small compared with the mass, and we can omit the
zeorth component of the polarization vector of the $K^{*0}$.

Now we only need to calculate the width $\Gamma$ associated with the diagram in Fig. \ref{fig::diagramBtoDpR}. Right away we see that since
\begin{equation}
\sum_{\text{pol. of $K^*$}} \epsilon_{K^*}^i \epsilon_{K^*}^j = \delta_{ij},
\end{equation}
\begin{widetext}
Eq. \eqref{eq:t_total} reduces to
\begin{equation}
 t_{B^- \rightarrow D^* \pi R} =g_{K^- K^+, R}  \ g C i\int \frac{d^4 q}{(2 \pi)^4} \ \frac{\vec{\epsilon}_{D^*} \cdot (\vec{\tilde{p}}_{K^+}-\vec{\tilde{p}}_{\pi})}{q^2-m_K^2+i \epsilon} \frac{1}{(P-q)^2-m^2_{K^*}+i \epsilon} \frac{1}{(P-q-k)^2 - m_K^2+i \epsilon},\label{eq_amp1}
\end{equation}
\end{widetext}
where $\vec{\tilde{p}}_{K^+}= \vec{P}-\vec{q}-\vec{k}= -(\vec{q}+\vec{k})$ and $\vec{\tilde{p}}_{\pi}= \vec{k}$.\\
Defining $f(\vec{q},\vec{k})$ as a product of the three propagators in
Eq.~(\ref{eq_amp1}), we can use the formula,
\begin{equation*}
\int d^3 q q_i f(\vec{q}, \vec{k}) = k_i \int d^3 q \frac{\vec{q} \cdot \vec{k}}{|\vec{k}|^2} f( \vec{q}, \vec{k}),
\end{equation*}
which follows from the fact that the $\vec{k}$ is the only vector
not integrated in the integrand of Eq.~(\ref{eq_amp1}).
Then, Eq.~(\ref{eq_amp1}) becomes
\begin{equation}
\label{eq:t_total2}
t_{B^- \rightarrow D^* \pi R} = -\vec{\epsilon}_{D^*} \cdot \vec{k} \ g_{K^- K^+, R} \ g C \  t_T,
\end{equation}
with
\begin{widetext}
\begin{equation}
t_T=i \int \frac{d^4 q}{(2 \pi)^4}  \ \left(2+ \frac{\vec{q} \cdot \vec{k}}{|\vec{k}|^2}\right) \frac{1}{q^2-m_K^2+i \epsilon} \frac{1}{(P-q)^2-m^2_{K^*}+i \epsilon} \frac{1}{(P-q-k)^2 - m_K^2+i \epsilon}.\label{eq_tT}
\end{equation}
\end{widetext}
Squaring and summing over the polarizations of $D^*$, Eq. \eqref{eq:t_total2} becomes
\begin{equation}
\label{eq:integrate}
 \sum_{\rm pol}|t_{B^- \rightarrow D^* \pi R}|^2 =|\vec{k}|^2 \ g^2_{K^- K^+, R} \ g^2 C^2  |t_T|^2,
\end{equation}

As given in Ref.~\cite{Bayar:2016ftu}, the analytical integration of
$t_T$ in Eq.~(\ref{eq_tT}) over $q^0$ leads to
\begin{widetext}
\begin{align}
 t_T =& \int \frac{d^3 q}{(2 \pi)^3}   \left(2+ \frac{\vec{q} \cdot \vec{k}}{|\vec{k}|^2}\right) \frac{1}{8 \omega^* \omega \omega'} \frac{1}{k^0-\omega'-\omega^*} \frac{1}{P^0+\omega+\omega'-k^0} \frac{1}{P^0-\omega-\omega'-k^0 + i \epsilon} \times\notag\\
\label{eq:tsing_int}
&\times \frac{\{2P^0 \omega + 2 k^0 \omega' -2[\omega+\omega'][\omega+\omega'+\omega^*] \}}{P^0-\omega^*-\omega+i\epsilon},
\end{align}
\end{widetext}
with $\omega^*(\vec{q})=\sqrt{m^2_{K^{* 0}} +|\vec{q}|^2}$, $\omega'(\vec{q})=\sqrt{m^2_{K} +|\vec{q}+\vec{k}|^2}$ and  $\omega(\vec{q})=\sqrt{m_{K}^2 +|\vec{q}|^2}$. To regularize the integral in Eq.~\eqref{eq:tsing_int} we use the same cutoff of the meson loop that will be used to calculate $t_{K^+ K^- \rightarrow \pi^ 0 \eta}$ and $t_{K^+ K^-
\rightarrow\pi^+ \pi^-}$ (Eq.~\eqref{eq:LSeq}), $\theta(q_{\text{max}}-|q^*|)$, where $\vec{q}^{\ *}$ is the $K^-$ momentum in the $R$ rest frame \cite{Bayar:2016ftu}.

In in Ref.~\cite{Bayar:2016ftu} it was found that there is a singularity associated with this type of loop functions when Eq. (18) of Ref.~\cite{Bayar:2016ftu} is satisfied. From that equation we can determine that the singularity will appear around $M_{\text{inv}}(\pi R) = 1418 \ \text{MeV}$.

To be completely correct in our analysis we have to use the width of $K^{* 0}$. We implement that replacing $\omega^* \rightarrow \omega^* - i \frac{\Gamma_{K^{*}}}{2}$ in Eq. \eqref{eq:tsing_int}, which will reduce the singularity to a peak~\cite{Bayar:2016ftu}.

For the three body decay of $B^-
\rightarrow D^{* 0} \pi^- R$
in Fig. \ref{fig::diagramBtoDpR}, the mass distribution is given by
\begin{equation}
\label{eq:totalwidth1}
\frac{d \Gamma}{d M_{\text{inv}}(\pi R)}= \frac{1}{(2 \pi)^3}
\frac{|\vec{p}_{D^{*}}| |\vec{\tilde{p}}_{\pi}|}{4 M_B^2}  \sum_{\rm pol.}\left|t_{B^- \rightarrow D^* \pi R} \right|^2,
\end{equation}
with
\begin{subequations}
\begin{align}
&	|\vec{p}_{D^{*}}| = \frac{\lambda^{1/2} (M_B^2, m^2_{D^{*}}, M^2_{\text{inv}}(\pi R))}{2 M_B},	\\
\label{eq::Kformula}
&   |\vec{\tilde{p}}_{\pi}|=|\vec{k}|=\frac{\lambda^{1/2} (M^2_{\text{inv}}(\pi R), m^2_{\pi}, M^2_R)}{2 M_{\text{inv}}(\pi R)}
\end{align}
\end{subequations}
With Eq.~(\ref{eq:t_total2}) and a factor
$1/\Gamma_{B^-}$, the mass distribution of $B^-$ decaying into $D^* \pi
R$ is written as
\begin{widetext}
\begin{equation}
\label{eq:final_totalW}
\frac{1}{\Gamma_{B^-}}\frac{d \Gamma}{d M_{\text{inv}}(\pi R)} = \frac{C^2}{\Gamma_{B^-}}\frac{g^2}{(2 \pi)^3} \frac{|\vec{p}_{D^{*}}| |\vec{k}|}{4 M_B^2} \  |\vec{k}|^2\cdot\left| t_T \times \ g_{K^- K^+, R} \right|^2,
\end{equation}
\end{widetext}
where $\frac{C^2}{\Gamma_{B^-}}$ is given in Eq.~\eqref{eq:const}.

However, the problem here is that the $a_0$ and $f_0$ are not stable
particles, but resonances that have a width and decay to $\pi^0 \eta$
and $\pi^+ \pi^-$, respectively.
To solve this without having to consider $R$ a virtual particle and
having a four body decay, we can consider the resonance as a normal
particle but we add a mass distribution to the decay width in
Eq. \eqref{eq:totalwidth1},
\begin{widetext} 
\begin{equation}
\label{eq:totalwidth_new}
\frac{d \Gamma}{d M_{\text{inv}}(\pi R)}=\frac{1}{(2 \pi)^3} \int d M^2_{\text{inv}}(R) \ (-\frac{1}{\pi}\text{Im}D )\  |g_{K^- K^+, R}|^2 \frac{|\vec{p}_{D^{*}}| |\vec{\tilde{p}}_{\pi}|}{4 M_B^2} \ \overline{\sum} \sum \left|\tilde{t}_{B^- , D^* \pi R} \right|^2,
\end{equation}
\end{widetext} 
with 
\begin{equation}
D= \frac{1}{M_{\text{inv}}^2(R) - M_R^2 + i M_R \Gamma_R},
\end{equation}
where $M_{\text{inv}}(R)$ stands for $M_{\text{inv}}(\pi^0  \eta)$ and $M_{\text{inv}}(\pi^+ \pi^-)$ for $R=a_0$ and $f_0$, respectively, and $\tilde{t}_{B^- , D^* \pi R}  = t_{B^- \rightarrow D^* \pi R}/g_{K^- K^+, R} $.
What Eq.~\eqref{eq:totalwidth_new} is accomplishing is a convolution of Eq.~\eqref{eq:totalwidth1} with the mass distribution of the $R$ resonance given by its spectral function.

Notice also that in the limit of $\Gamma_R \rightarrow 0$, $ i \text{Im} D = -i \pi \delta (M_{\text{inv}}^2(R)- M_R^2)$ and we recover Eq. \eqref{eq:totalwidth1}.
Evaluating explicitly the imaginary part of $D$, Eq. \eqref{eq:totalwidth_new} becomes
\begin{widetext}
\begin{equation}
\label{eq:totalwidth_new2}
\frac{d \Gamma}{d M_{\text{inv}}(\pi R)}= \frac{1}{(2 \pi)^3} \int d M^2_{\text{inv}}(R) \ \frac{1}{\pi}  \  |g_{K^- K^+, R}|^2 \frac{|\vec{p}_{D^{*}}| |\vec{\tilde{p}}_{\pi}|}{4 M_B^2} \ \overline{\sum} \sum \left|\tilde{t}_{B^- , D^* \pi R} \right|^2 \frac{M_R \Gamma_R}{(M_{\text{inv}}^2(R) - M_R^2)^2+(M_R \Gamma_R)^2}.
\end{equation}
\end{widetext} 

Now, for the case of $a_0(980)$, we only have the decay $a_0 \rightarrow \pi^0 \eta$ (we neglect the small $K \bar{K}$ decay fraction), and thus,
\begin{equation}
\Gamma_{a_0}= \frac{1}{8 \pi} \frac{|g_{a_0 \rightarrow \pi^0 \eta}|^2}{M^2_{\text{inv}}(\pi^0\eta)} |\vec{\tilde{q}}_{\eta}|,
\end{equation}
with
\begin{equation}
|\vec{\tilde{q}}_{\eta}| = \frac{\lambda^{1/2}(M_{\text{inv}}^2(\pi^0\eta), m_{\pi}^2, m_{\eta}^2)}{2 M_{\text{inv}}(\pi^0\eta)}.
\end{equation}
Then Eq. \eqref{eq:totalwidth_new2} becomes
\begin{widetext}
\begin{equation}
\label{eq:totalwidth_new3}
\frac{d \Gamma}{d M_{\text{inv}}(\pi a_0)}= \frac{1}{(2 \pi)^3}\ \int d M^2_{\text{inv}}(\pi^0\eta) \frac{|\vec{p}_{D^{*}}| |\vec{\tilde{p}}_{\pi}|}{4 M_B^2}  \ \overline{\sum} \sum \left|\tilde{t}_{B^- , D^* \pi R} \right|^2  \frac{M_{a_0}  |g_{a_0 \rightarrow \pi^0 \eta}|^2 |g_{K^- K^+ \rightarrow a_0}|^2}{(M_{\text{inv}}^2(\pi^0\eta) - M_{a_0}^2)^2+(M_{a_0} \Gamma_{a_0})^2}\frac{1}{8 \pi^2} \frac{|\vec{\tilde{q}}_{\eta}|}{M^2_{\text{inv}}(\pi^0\eta)}.
\end{equation}
\end{widetext} 
But since for the resonance we have formally,
\begin{equation}
\frac {|g_{a_0 \rightarrow \pi^0 \eta}|^2 |g_{K^- K^+ \rightarrow a_0}|^2}{(M_{\text{inv}}^2(\pi^0\eta) - M_{a_0}^2)^2+(M_{a_0} \Gamma_{a_0})^2} = \left|t_{K^+ K^- \rightarrow \pi^0 \eta} \right|^2,
\end{equation}
Eq. \eqref{eq:totalwidth_new3} reduces to
\begin{widetext}
\begin{equation}
\label{eq:totalwidth_new4}
\frac{d^2 \Gamma}{d M_{\text{inv}}(\pi a_0) d M_{\text{inv}}(\pi^0\eta)} = \frac{1}{(2 \pi)^5} \frac{|\vec{p}_{D^{*}}| |\vec{k}| |\vec{\tilde{q}}_{\eta}|}{4 M_B^2} \ \overline{\sum} \sum \left|\tilde{t}_{B^- , D^* \pi R} \times  t_{K^+ K^- \rightarrow \pi^0 \eta}\right|^2,
\end{equation}
\end{widetext}
where we approximated $\minv(\pi^0\eta)$ as $M_R$.
For the case of $f_0(980)$, $f_0 \rightarrow \pi^+ \pi^-$ is not the only possible decay and as such $\Gamma_{f_0 \rightarrow \pi^+ \pi^-}$ will not be the same as the $\Gamma_R$ in Eq. \eqref{eq:totalwidth_new}. However, when we put $ \left|t_{K^+ K^- \rightarrow \pi^+ \pi^-} \right|^2$ in the end, we already select the $\pi \pi$ part of the $f_0$ decay. Thus, for the case of $f_0$ we just need to substitute, in Eq. \eqref{eq:totalwidth_new4}, $t_{K^+ K^- \rightarrow \pi^0 \eta} \rightarrow t_{K^+ K^- \rightarrow \pi^+ \pi^-}$, $M_{\text{inv}}(\pi a_0) \rightarrow M_{\text{inv}}(\pi f_0)$, $M_{\text{inv}}(\pi^0\eta) \rightarrow M_{\text{inv}}(\pi^+\pi^-)$, and $|\vec{\tilde{q}}_{\eta}| \rightarrow |\vec{\tilde{q}}_{\pi}|$, with
\begin{equation}
|\vec{\tilde{q}}_{\pi}| = \frac{\lambda^{1/2}(M_{\text{inv}}^2(\pi^+\pi^-), m_{\pi^+}^2, m_{\pi^-}^2)}{2 M_{\text{inv}}(\pi^+\pi^-)}.
\end{equation}
The amplitudes $t_{K^+ K^- \rightarrow \pi^ 0 \eta}$ and $t_{K^+ K^-
\rightarrow\pi^+ \pi^-}$ themselves are calculated based on the chiral
unitary approach, where the $a_0$ and $f_0$ appear as dynamically
generated states \cite{Oller:1997ti,Oller:1998hw}. The cutoff
parameter $q_{\text{max}}$ which appears for the regularization of the
meson loop function in the Bethe-Salpeter equation,
\begin{equation}
\label{eq:LSeq}
t=[1-V G]^{-1} V,
\end{equation}
is determined as  $q_{\text{max}}=600 \ \text{MeV}$ for the reproduction of the $a_0$ and $f_0$ peaks (around $980 \ \text{MeV}$ in invariant mass of $\pi^0 \eta$ or $\pi^+ \pi^-$)~\cite{a0amp,f0amp}.
In Eq. \eqref{eq:LSeq}, $t$, $V$, and $G$ are the meson amplitude, interaction kernel, and meson loop function, respectively.

Finally, we can substitute everything we have calculated into Eq. \eqref{eq:totalwidth_new4} and obtain,
\begin{widetext}
\begin{equation}
\label{eq:final_totalW2}
\frac{1}{\Gamma_{B^-}}\frac{d^2 \Gamma}{d M_{\text{inv}}(\pi R) d M_{\text{inv}}(R)} = \frac{g^2}{(2 \pi)^5} \frac{|\vec{p}_{D^{*}}| |\vec{\tilde{q}}_{\eta}| |\vec{k}|^3}{4 M_B^2} \  \left| t_T \times t_{K^+ K^- \rightarrow \pi^0 \eta (\pi^+ \pi^-)}\right|^2 \frac{C^2}{\Gamma_{B^-}}.
\end{equation}
\end{widetext}

\section{Results}

\begin{figure}[h!]
  \centering
  \includegraphics[scale = 0.55]{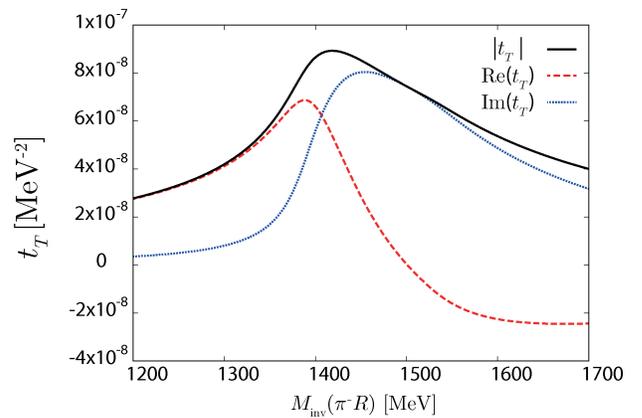}
  \captionsetup{justification=raggedright}
  \caption{Triangle amplitude $t_T$ for the decay $B^- \rightarrow D^{*
 0}\pi R$.
 {We take $\minv(R)=$ 980 MeV.}}
  \label{fig::loopDstar}
\end{figure}

Let us begin by showing in Fig. \ref{fig::loopDstar} the contribution of
the triangle loop (defined in Eq. \eqref{eq:tsing_int}) to the total
amplitude.
We plot the real and imaginary parts of $t_T$, as well as the
absolute value with $\minv(R)$ fixed at 980 MeV.
As can be observed, there is a peak around $1420 \
\text{MeV}$, as predicted by Eq. (18) of Ref.~\cite{Bayar:2016ftu}.

\begin{figure}[h!]
  \centering
  \includegraphics[scale = 0.55]{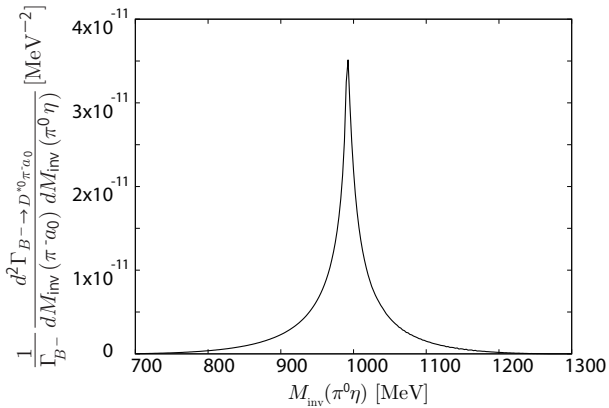}
  \captionsetup{justification=raggedright}
  \caption{The derivative of the mass distribution of $B^- \rightarrow D^{* 0} \pi^- \pi^0 \eta $ with regards to $M_{\text{inv}}(a_0)$.}
  \label{fig::Dstarpieta}
\end{figure}

\begin{figure}[h!]
  \centering
  \includegraphics[scale = 0.55]{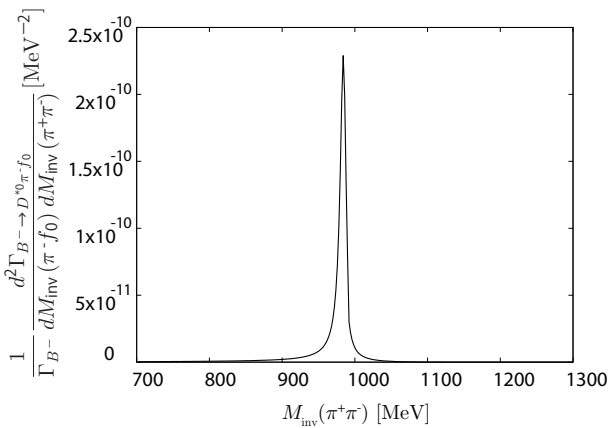}
  \captionsetup{justification=raggedright}
  \caption{The derivative of the mass distribution of $B^- \rightarrow D^{* 0} \pi^- \pi^+ \pi^- $ with regards to $M_{\text{inv}}(f_0)$.}
  \label{fig::Dstarpipi}
\end{figure}

In Figs.~\ref{fig::Dstarpieta} and \ref{fig::Dstarpipi} we plot
Eq. \eqref{eq:final_totalW2} for both $B^- \rightarrow D^{* 0} \pi^- \eta
\pi^0$ and $B^- \rightarrow D^{* 0} \pi^- \pi^+ \pi^-$, respectively, by
fixing $M_{\text{inv}}(\pi R)=1418 \ \text{MeV}$,
which is the position of the triangle singularity, and varying
$M_{\text{inv}}(R)$. We can see a strong peak around $980 \ \text{MeV}$
and consequently we see that most of the contribution to our width
$\Gamma$ will come from $M_{\text{inv}}(R)=M_R$. For
Fig. \ref{fig::Dstarpieta} the dispersion is bigger, we have strong
contributions for $M_{\text{inv}}(\pi^0 \eta) \in [880,1080]$. However, for
Fig. \ref{fig::Dstarpipi} most of the contribution comes from
$M_{\text{inv}}(\pi^+ \pi^-) \in [940,1020]$. The conclusion is that when we
calculate the mass distribution $\frac{d \Gamma}{d M_{\text{inv}}(\pi a_0)}$, we can restrict the integral in
$M_{\text{inv}}(R)$ to the limits already mentioned. 

\begin{figure}[h!]
  \centering
  \includegraphics[scale = 0.55]{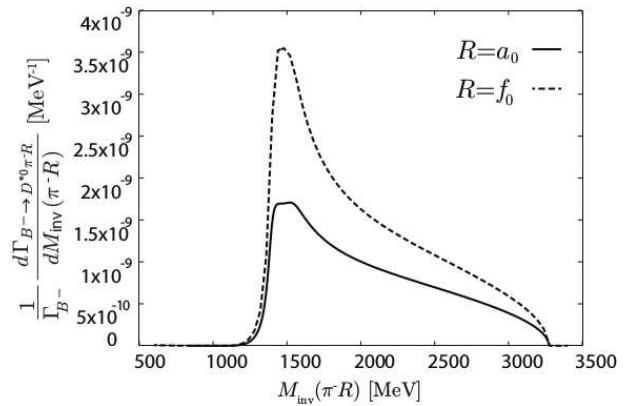}
  \captionsetup{justification=raggedright}
  \caption{The mass distribution of $B^- \rightarrow D^{* 0} \pi^- \pi^0 \eta$ (full line) and  $B^- \rightarrow D^{* 0} \pi^- \pi^+ \pi^- $ (dashed line).}
  \label{fig::DstarR_mass}
\end{figure}

When we integrate over $M_{\text{inv}}(R)$ we obtain $\frac{d \Gamma}{d M_{\text{inv}}(\pi R)}$ which we show in Fig.~\ref{fig::DstarR_mass}.
We see a clear peak of the distribution around $1420 \ \text{MeV}$, for $f_0$ and $a_0$ production. However, we also see that the distribution stretches up to large values of $M_{\text{inv}}(\pi R)$ where the phase space of the reaction finishes. This is due to the $|\vec{k}|^3$ factor in Eq.~\eqref{eq:final_totalW2} that contains a $|\vec{k}|$ factor from phase space and a $|\vec{k}|^2$ factor from the dynamics of the process, as we can see in Eq.~\eqref{eq:integrate}. Yet, a clear peak in $M_{\text{inv}}(\pi^- R)$ can be seen for both the $B^- \rightarrow D^{* 0} \pi^- f_0$ and $B^- \rightarrow D^{* 0} \pi^- a_0$ reactions.

Integrating now $\frac{d \Gamma}{d M_{\text{inv}}(\pi a_0)}$ and $\frac{d \Gamma}{d M_{\text{inv}}(\pi f_0)}$ over the $M_{\text{inv}}(\pi a_0)$ ($M_{\text{inv}}(\pi f_0)$) masses in Fig.~\ref{fig::DstarR_mass}, we obtain the branching fractions
\begin{subequations}
\begin{align}
& \br(B^-\rightarrow D^{*0}\pi^- a_0; a_0 \rightarrow \pi^0\eta) =  1.66 \times 10^{-6} ,\\
& \br(B^-\rightarrow D^{*0}\pi^- f_0; f_0 \rightarrow \pi^+\pi^-)=  2.82 \times 10^{-6}.
\end{align}
\end{subequations}
These numbers are within measurable range.

Note that we have assumed all the strength of $\pi^0 \eta$ from $880 \ \text{MeV}$ to $1080 \ \text{MeV}$ to be part of the $a_0$ production, but in an experimental analysis one might associate part of this strength to a background. We note this in order to make proper comparison with these results when the experiment is performed.

The shape of $t_T$ in Fig.~\ref{fig::loopDstar} requires some extra comment. We see that Im$(t_T)$ peaks around $1420$ MeV, where the triangle singularity is expected. However Re$(t_T)$ also has a peak around $1390$ MeV. This picture is not standard. Indeed, in Ref.~\cite{daris}, where a triangle singularity is disclosed for the process $N (1835) \rightarrow \pi N (1535)$, $t_T$ has the real part peaking at the place of the triangle singularity and Im$(t_T)$ has no peak. In Ref.~\cite{debastiani22}, a triangle singularity develops in the $\gamma p \rightarrow p \pi^0 \eta \rightarrow \pi^0 N (1535)$ process and there Im$(t_T)$ has a peak at the expected energy of the triangle singularity while the Re$(t_T)$ has no peak. Similarly, in the study of $N (1700) \rightarrow \pi \Delta$ in Ref.~\cite{roca} a triangle singularity develops and here Im$(t_T)$ has a peak but Re$(t_T)$ has not. However, the double peak in the real and imaginary parts of $t_T$ is also present in the study of the $B^- \rightarrow K^- \pi D_{s0}^+$ reaction in Ref.~\cite{Sakai:2017hpg}. This latter work has a loop with $D^0 K^{* 0} K^+$, and by taking $\Gamma_{K^*} \rightarrow 0$, $\epsilon \rightarrow 0$ the peak of Im$(t_T)$ was identified with the triangle singularity while the peak in the Re$(t_T)$ was shown to come from the threshold of $D^0 K^{* 0}$. In the present case the situation is similar: The peak of Im$(t_T)$ at about $1420$ MeV comes from the triangle singularity while the one just below $1400$ MeV comes from the threshold of $K^{* 0} K^-$ in the diagram of Fig.~\ref{fig::diagramBtoDpR}, which appears at $1386$ MeV. Yet, by looking at $|t_T|$ in Fig.~\ref{fig::loopDstar} and the region of the peak of $\frac{d \Gamma}{d M_{\text{inv}}}$ in Fig.~\ref{fig::DstarR_mass}, we can see that this latter peak comes mostly from the triangle singularity.

\section{Summary}

We have performed the calculations for the reactions $B^- \rightarrow D^{* 0} \pi^- a_0(980); a_0 \rightarrow \pi^0 \eta$ and $B^- \rightarrow D^{* 0} \pi^- f_0(980); f_0 \rightarrow \pi^+ \pi^-$. The starting point is the reaction $B^- \rightarrow D^{* 0} K^{* 0} K^-$, which is a Cabibbo favored process and for which the rates are tabulated in the PDG~\cite{pdg} and are relatively large. Then we allow the $K^{* 0}$ to decay into $\pi^- K^+$ and the $K^+ K^-$ fuse to give the $f_0(980)$ or the $a_0(980)$. Both of them are allowed, since the $K^{* 0} K^-$ state does not have a particular isospin. The triangle diagram corresponding to this mechanism develops a triangle singularity at about $1420 \ \text{MeV}$ in the invariant masses of $\pi^- f_0$ or $\pi^- a_0$, and makes the process studied relatively large, having a prominent peak in those invariant mass distributions around $1420 \ \text{MeV}$.

We evaluate $\frac{d^2 \Gamma}{d M_{\text{inv}}(\pi^- a_0) d M_{\text{inv}}(\pi^0 \eta)}$, and $\frac{d^2 \Gamma}{d M_{\text{inv}}(\pi^- f_0) d M_{\text{inv}}(\pi^+ \pi^-)}$ and see clear peaks in the $M_{\text{inv}}(\pi^0 \eta)$, $M_{\text{inv}}(\pi^+ \pi^-)$ distributions, showing clearly the $a_0(980)$ and $f_0(980)$ shapes. Integrating over $M_{\text{inv}}(\pi^0 \eta)$ and $M_{\text{inv}}(\pi^+ \pi^-)$ we obtain $\frac{d \Gamma}{d M_{\text{inv}}(\pi a_0)}$ and $\frac{d \Gamma}{d M_{\text{inv}}(\pi f_0)}$ respectively, and these distributions show a clear peak for $M_{\text{inv}}(\pi a_0)$, $M_{\text{inv}}(\pi f_0)$ around $1420 \ \text{MeV}$. This peak is a consequence of the triangle singularity, and in this sense the work done here should be a warning not to claim a new resonance when this peak is seen in a future experiment. On the other hand, the results make predictions for an interesting effect of a triangle singularity in an experiment that is feasible in present experimental facilities. The rates obtained are also within measurable range. Finding new cases of triangle singularities is of importance also, because their study will give incentives to update present analysis tools to take into account such possibility when peaks are observed experimentally, avoiding the natural tendency to associate those peaks to resonances.

\section*{Acknowledgements}
R.P.~Pavao wishes to thank the Generalitat Valenciana in the program
Santiago Grisolia.
This work is partly supported by the Spanish Ministerio de Economia y
Competitividad and European FEDER funds under the contract number FIS2014-57026-REDT,
FIS2014-51948-C2-1-P, and FIS2014-51948-C2-2-P, and the Generalitat
Valenciana in the program Prometeo II-2014/068.

\end{document}